\documentclass{article}
\usepackage{spconf,amsmath,amssymb,graphicx,hyperref}
\usepackage{booktabs}
\usepackage{multirow}
\usepackage{array}
\usepackage[dvipsnames]{xcolor}
\usepackage{url}

\title{Prime-EVC: One-Step Expressive Emotional Voice Conversion via Emotion-Primed Flow Matching}

\name{\begin{tabular}{@{}c@{}}
  Yu Pan$^{1\dagger}$\thanks{$\dagger$~Corresponding author.}, Xiaojuan Cai$^{1}$, Liang Zhang$^{2}$, Yang Hou$^{4}$, Tianjiao Zhang$^{5}$, Yuguang Yang$^{3}$, Jixun Yao$^{3}$, \\ Hiroshi Koide$^{1}$,
  Lei Ma$^{6}$, Jianjun Zhao$^{1}$
\end{tabular}}
\address{
  $^{1}$Kyushu University, Japan \quad $^{2}$Donghua University, China \quad $^{3}$Tencent Hunyuan, China \\
  $^{4}$National Institute of Informatics, Japan \quad $^{5}$Independent Researcher \quad $^{6}$The University of Tokyo, Japan}

\begin{document}
\maketitle

\begin{abstract}
Recent diffusion- and conditional flow matching (CFM)-based emotional voice conversion (EVC) systems achieve strong conversion quality. However, they typically rely on multiple sampling steps, because the conditioning specifies which emotion to impose while its acoustic realization is left to the sampler. In this paper, we propose \mbox{Prime-EVC}, a one-step expressive EVC architecture based on emotion-primed flow matching. Our key idea is to reallocate the emotional conversion between the conditioning and the sampler: a FiLM-fused speaker-emotion style, an emotional variance adaptor, and direct emotion and spectro-temporal supervision assemble an emotion-rich conditioning upstream, rendering the residual conversion left to the CFM decoder attainable within one Euler step. Experiments on the ESD corpus demonstrate that single-step \mbox{Prime-EVC} surpasses five representative EVC systems on every quality metric and exceeds $4$- and $100$-step DurFlex-EVC in subjective naturalness and emotion while running $3.4$--$50\times$ faster than both.
\end{abstract}

\section{Introduction}
\label{sec:intro}

Emotional voice conversion (EVC) aims to transform the emotional state of an utterance, a para-linguistic attribute, into a target emotion while preserving its linguistic content and non-linguistic speaker identity~\cite{ref:evc}. 

Early EVC builds on generative adversarial networks (GANs)~\cite{ref:stargan,ref:gan} or sequence-to-sequence models~\cite{ref:seq2seq,zhou21b_interspeech}, while later works broaden the repertoire via intensity control~\cite{ref:einet,ref:emovox} and textless formulations that better decouple emotion from timbre and content~\cite{ref:srcfilter,ref:textless}. While efficient, these objectives are prone to over-smoothing, leaving the converted speech weakly or unnaturally emotional. 
To address this, recent studies adopt diffusion- or flow-matching-based decoders~\cite{ref:durflex,su2025diffemotionvc,ref:clapfm}. DurFlex-EVC~\cite{ref:durflex} pairs discrete-unit content modeling and style conditioning with a diffusion decoder, attaining strong performance but at the cost of many-step iterative denoising. 
Conditional flow matching (CFM)~\cite{ref:cfm,ref:matcha} decodes in fewer ordinary-differential-equation (ODE) steps, i.e., a smaller number of function evaluations (NFE), and is thus a natural replacement. ClapFM-EVC~\cite{ref:clapfm} first validates CFM for EVC via CLAP \cite{elizalde2023clap,pan2024gemo}-based emotion control, yet operates any-to-one and still requires multiple ODE evaluations. Efforts to accelerate such generation distill a multi-step teacher~\cite{reflow} or train a one-step generator~\cite{ma2026meanvc}, but all target efficiency rather than emotion conversion.

In this paper, we contend that leaving the acoustic realization of the target emotion to the sampler is what obstructs low-NFE EVC. Emotion is carried by fine-grained acoustic variation, which a single deterministic ODE step smooths toward the conditional mean, and more steps sharpen the sample without supplying what the conditioning leaves unspecified (Sec.~\ref{sec:nfe}).
We therefore present \textbf{Prime-EVC}, which \emph{offloads} emotion modeling onto a directly supervised, emotion-rich conditioning, so that the mean a single Euler step settles on is already expressive.
To that end, an emotional variance adaptor supplies the target-emotion pitch and energy, a FiLM-fused style accounts for the speaker-dependent form the emotion takes, and emotion and spectro-temporal losses supervise the conditioning directly. 
A Diffusion-Transformer~\cite{ref:dit} (DiT)-based CFM decoder then recovers the target mel-spectrogram features, vocoded by BigVGAN~\cite{ref:bigvgan}, atop a text-free content/duration backbone~\cite{ref:durflex}. 
Extensive experiments on ESD~\cite{ref:esd} dataset demonstrate that, with a single ODE step, Prime-EVC outperforms five representative EVC systems in objective quality and exceeds the best multi-step baseline in subjective ratings.

\begin{figure*}[htbp]
\centering
    \includegraphics[height=4.8cm,width=!]{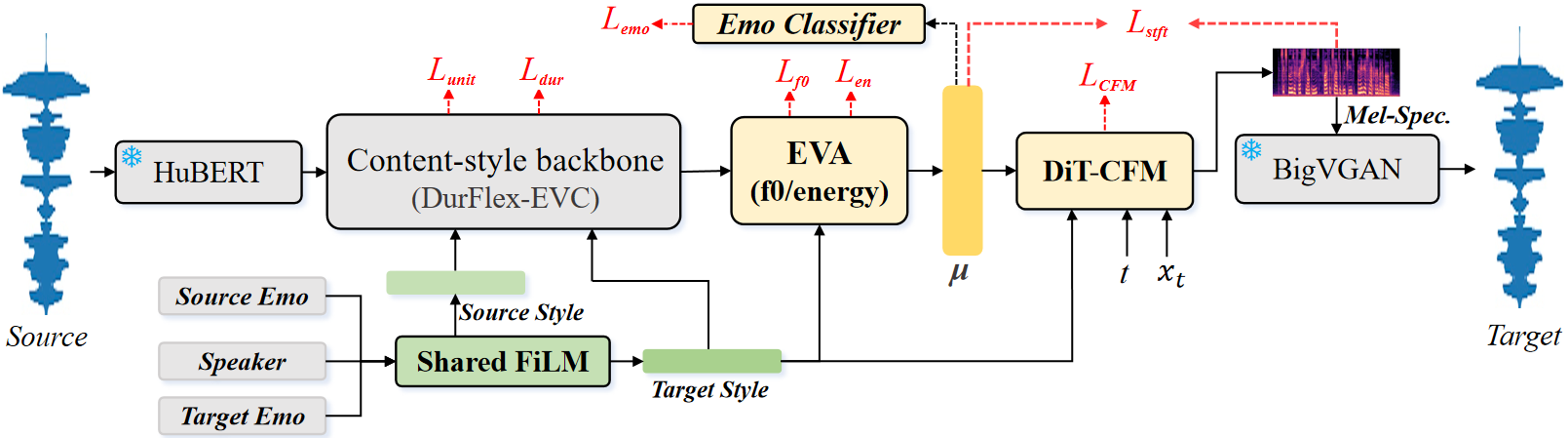}
    \caption{Overview of Prime-EVC, an \emph{encode--condition--render} pipeline.}
    \label{fig:arch}
\end{figure*}

\section{Proposed Method}
\label{sec:method}

\subsection{Overview}
\label{sec:overview}
As shown in Fig.~\ref{fig:arch}, Prime-EVC follows an \emph{encode--condition--render} pipeline. The \emph{encode} stage disentangles speech into a speaker- and emotion-agnostic content sequence, via a frozen HuBERT encoder~\cite{ref:hubert} and a duration predictor, following DurFlex-EVC~\cite{ref:durflex}. 
The \emph{condition} stage assembles the emotion-rich \emph{conditioning} $\mu$ from a fused speaker--emotion style and frame-level target-emotion prosody, under direct supervision.
The \emph{render} stage employs a DiT-CFM decoder conditioned on the frame-level $\mu$ and the global style $c$ to produce the target mel, followed by BigVGAN.

\subsection{Style-Aware Emotion Conditioning}
\label{sec:style}
Emotion is inherently \emph{speaker-dependent}, as the same target emotion surfaces through different spectral and prosodic signatures across voices~\cite{pan2023msac,pan2024gmp}. We therefore fuse the speaker code $s$ and a learnable $5$-class emotion lookup $e$ via feature-wise linear modulation (FiLM)~\cite{ref:film,pan2026s2st}:
\begin{equation}
c=\gamma(e)\odot s+\beta(e),
\end{equation}
where $c$ denotes the fused style, $\gamma(\cdot)$ and $\beta(\cdot)$ are emotion-conditioned scale and shift functions. The multiplicative term $\gamma(e)\odot s$ lets the target emotion \emph{re-weight} each speaker dimension while $\beta(e)$ relocates it. Applied to the source and target emotions, it yields source and target styles that de-stylize and re-stylize the content. Training uses source$=$target (reconstruction), while inference uses source$\neq$target (conversion).
To regularize this style, we scale the emotion embedding by a stochastic intensity factor $\alpha\sim\mathcal{U}(0.2,1.2)$, mildly extrapolating beyond full intensity $\alpha{=}1$, before fusion during training, fixing $\alpha{=}1$ at inference. Since $e$ encodes an emotion as a \emph{direction} in embedding space, perturbing its magnitude exposes the model to a continuum of emotion strengths and encourages correct conversion across that range.

\subsection{Emotional Variance Adaptation}
\label{sec:pim}
While $c$ fixes \emph{which} emotion to impose, emotion is heard through prosody, and nothing so far predicts the pitch and energy the target emotion calls for, leaving the decoder to invent them while sampling. 
We therefore relocate this generation upstream into the conditioning: a FastSpeech-style~\cite{ref:fastspeech2} \emph{emotional variance adaptor} (EVA) predicts frame-level target-emotion log-pitch $\hat{p}$ and energy $\hat{n}$ from $z_{\mathrm{frame}}$ conditioned on $c$, and embeds them back into content stream:
\begin{equation}
(\hat{p},\hat{n})=\mathrm{EVA}(z_{\mathrm{frame}},c),
\end{equation}
\begin{equation}
\tilde{z}_{\mathrm{frame}}=z_{\mathrm{frame}}+E_{p}(p^{\star})+E_{n}(n^{\star})
\end{equation}
from which the $80$-dim, mel-aligned conditioning follows as $\mu=\mathrm{Proj}_{\mu}(\tilde{z}_{\mathrm{frame}})$, where $\mathrm{Proj}_{\mu}$ is a linear layer and $E_{p},E_{n}$ are learned $1$-D embeddings. The injected contours $(p^{\star},n^{\star})$ follow teacher forcing: ground-truth $(p,n)$ in training and self-predicted $(\hat{p},\hat{n})$ at inference. 
Moreover, the predictor is learned in parallel under $\mathcal{L}_{f_0}{=}\lVert\hat{p}-p\rVert_2^2$ and $\mathcal{L}_{\mathrm{en}}{=}\lVert\hat{n}-n\rVert_2^2$.

\begin{table*}[t]
\centering
\caption{Overall comparison on the ESD corpus. Best among the conversion systems in \textbf{bold}.}
\label{tab:main}
\renewcommand{\arraystretch}{0.8} 
\begin{tabular}{lcccccccc}
\toprule
Method & nMOS$\uparrow$ & eMOS$\uparrow$ & UTMOS$\uparrow$ & CER$\downarrow$ & WER$\downarrow$ & EECS$\uparrow$ & SECS$\uparrow$ & RTF$\downarrow$ \\
\midrule
GT         &  4.02 $\pm 0.03$  &  4.06 $\pm 0.02$ & 3.65 & 2.15 & 5.69 & 0.96 & 0.84  & - \\
GT (vocoded)    &  3.96 $\pm 0.05$  & 4.01 $\pm 0.03$  & 3.42 & 2.39 & 5.85 & 0.95 & 0.84 & - \\
\midrule
StarGAN-EVC         & 2.05$\pm 0.06$    & 2.11 $\pm 0.04$    & 1.39 & 49.77 & 72.38 & 0.37 & 0.57 & \textbf{0.005}  \\
Seq2seq-EVC      & 2.72 $\pm 0.03$   & 2.69 $\pm 0.05$   & 1.48 & 22.19 & 37.62 & 0.39 & 0.61 & 0.233 \\
Emovox          & 3.06 $\pm 0.05$   & 3.17 $\pm 0.03$  & 2.11 & 18.69 & 34.18 & 0.49 & 0.69 & 0.211  \\
Textless-EVC         & 3.43 $\pm 0.06$   & 3.39 $\pm 0.04$  & 2.84 & 15.17 & 26.73 & 0.65 & 0.66 & 0.434 \\
DurFlex (NFE=1)     & 2.01 $\pm 0.05$   & 2.09 $\pm 0.06$  & 1.32 & 17.98 & 29.25 & 0.32 & 0.47 & 0.077  \\
DurFlex (NFE=4)     & 3.63 $\pm 0.05$   & 3.79 $\pm 0.05$  & 3.09 & 5.59 & 11.90 & 0.85 & 0.71 & 0.132  \\
DurFlex (NFE=100)  & 3.69 $\pm 0.04$   & 3.83 $\pm 0.03$  & 3.21 & 5.65 & 11.95 & 0.86 & 0.76 & 1.917 \\
\textbf{Prime-EVC (NFE=1)} &  \textbf{3.77 $\pm 0.03$}  & \textbf{3.90 $\pm 0.04$}  & \textbf{3.37} & \textbf{4.40} & \textbf{9.61} & \textbf{0.87} & \textbf{0.80} & 0.038  \\
\bottomrule
\end{tabular}
\end{table*}

\subsection{Conditioning-Level Supervision}
\label{sec:supervision}
To make the assembled $\mu$ an effective conditioning, we supervise it directly along two complementary aspects: emotional expressiveness and acoustic naturalness. A lightweight classifier $g_{\phi}$ on the pooled $\bar{\mu}{=}\frac1T\sum_t\mu_t$, trained by cross-entropy $\mathcal{L}_{\mathrm{emo}}$ against $y_e$, makes $\mu$ emotion-discriminative. 
Since the conditioning is consumed in a single evaluation and cannot be refined downstream, we regularize $\mu$'s temporal dynamics by matching, per mel channel, the magnitude of its \emph{modulation spectrum}~\cite{takamichi2016postfilters}, a classical remedy for spectral over-smoothing: each band's Hann-windowed trajectory is matched to that of the target mel $x_1$ through spectral-convergence ($\mathcal{L}_{\mathrm{sc}}$) and log-magnitude ($\mathcal{L}_{\mathrm{mag}}$) terms~\cite{yamamoto2020parallel}:
\begin{equation}
\mathcal{L}_{\mathrm{stft}}=\mathcal{L}_{\mathrm{sc}}(\mu,x_1)+\mathcal{L}_{\mathrm{mag}}(\mu,x_1).
\end{equation}
This gives $\mu$ a coarse spectro-temporal envelope rather than the pointwise copy an $\ell_1$ mel loss would enforce.

\subsection{DiT-based CFM Decoder}
\label{sec:dec}
Given the assembled conditioning $\mu$, the decoder converts the target mel-spectrogram via optimal-transport-based conditional flow matching (OT-CFM)~\cite{ref:cfm,ref:matcha}. For $z\sim\mathcal{N}(0,I)$, target $x_1$, and $\sigma_{\min}{=}10^{-6}$, the optimal-transport path and its target velocity field are:
\begin{equation}
x_t=\bigl(1-(1-\sigma_{\min})t\bigr)z+t\,x_1,\quad u=x_1-(1-\sigma_{\min})z.
\end{equation}
The DiT module $v_\theta(x_t,t;\mu,c)$ regresses this field: the interpolated mel $x_t$ is concatenated with $\mu$ along the feature axis and, with positional embeddings, forms the input tokens, while the CFM time $t$ and the global style $c$ are injected via adaptive layer normalization. 

\subsection{Training Objective}
\label{sec:train}
All modules of Prime-EVC are optimized jointly and end-to-end. The decoder is trained with the OT-CFM loss $\mathcal{L}_{\mathrm{CFM}}=\mathbb{E}_{t,z,x_1}\lVert v_\theta(x_t,t;\mu,c)-u\rVert_2^2$. The conditioning is shaped by the emotion and spectro-temporal objectives of Secs.~\ref{sec:pim}--\ref{sec:supervision}, and two backbone terms (the HuBERT-unit cross-entropy $\mathcal{L}_{\mathrm{unit}}$ and the stochastic-duration objective $\mathcal{L}_{\mathrm{dur}}$) inherited unchanged from ~\cite{ref:durflex}. Therefore, the overall objective of the proposed Prime-EVC is:
\begin{align}
\mathcal{L}=&~\mathcal{L}_{\mathrm{CFM}}+\lambda_{\mathrm{stft}}\mathcal{L}_{\mathrm{stft}}+\lambda_{\mathrm{unit}}\mathcal{L}_{\mathrm{unit}}+\lambda_{\mathrm{dur}}\mathcal{L}_{\mathrm{dur}}\nonumber\\
&+\lambda_{f_0}\mathcal{L}_{f_0}+\lambda_{\mathrm{en}}\mathcal{L}_{\mathrm{en}}+\lambda_{\mathrm{emo}}\mathcal{L}_{\mathrm{emo}},
\end{align}
where the auxiliary weights are empirically set to $\lambda_{\mathrm{stft}}{=}0.3$ with the remainder fixed at $0.1$.

\section{Experimental Setup}

\label{sec:setup}
\noindent\textbf{Dataset and protocol.} All experiments are conducted on the English portion of ESD~\cite{ref:esd}, a parallel corpus of five emotions (Neutral, Angry, Happy, Sad, Surprise). Using its official split and $16$\,kHz audio, and following DurFlex-EVC's protocol, each test utterance is converted to the four other emotions, yielding $6000$ conversions.

\noindent\textbf{Baselines.} We compare Prime-EVC against five representative EVC systems, including: the GAN-based StarGAN-EVC\footnote{\url{https://github.com/glam-imperial/EmotionalConversionStarGAN}}~\cite{ref:stargan}, the sequence-to-sequence Seq2seq-EVC\footnote{\url{https://github.com/KunZhou9646/seq2seq-EVC}}~\cite{ref:seq2seq}, the intensity-controllable Emovox\footnote{\url{https://github.com/KunZhou9646/Emovox}}~\cite{ref:emovox}, the textless Textless-EVC\footnote{\url{https://github.com/facebookresearch/fairseq/tree/main/examples/emotion_conversion}}~\cite{ref:textless}, and the diffusion-based DurFlex-EVC\footnote{\url{https://github.com/hs-oh-prml/DurFlexEVC}}~\cite{ref:durflex}. For a controlled comparison, all baselines are retrained on the same split with their official implementations. 

\noindent\textbf{Evaluation metrics.} For the subjective study, $16$ English-fluent listeners rate $100$ converted speech per system on two 5-point MOS scales, naturalness (nMOS) and emotion (eMOS), reported as means with $95\%$ confidence intervals. Objectively, we assess five axes: naturalness (UTMOS~\cite{ref:utmos}), intelligibility (CER/WER from Whisper-large-v3\footnote{\url{https://huggingface.co/openai/whisper-large-v3}}, with its English text normalizer applied to both hypothesis and reference), emotion conversion (EECS, emotion2vec\footnote{\url{https://huggingface.co/emotion2vec}} cosine similarity to a same-speaker target-emotion reference), speaker similarity (SECS, Resemblyzer cosine similarity to the same-speaker target-emotion reference), and inference speed (RTF). The mel-based systems (Textless-EVC, DurFlex-EVC, and Prime-EVC) share an identical BigVGAN retrained at 16,kHz on LibriTTS and VCTK datasets, so differences reflect the acoustic models. Its copy-synthesis topline is reported in Table~\ref{tab:main}, where the GT rows use a different same-speaker, same-emotion reference and thus mark each metric's empirical upper bound.

\noindent\textbf{Implementation details.} Prime-EVC comprises $\sim$98M parameters, fewer than DurFlex-EVC's $\sim$159M, and retains the HuBERT-unit
vocabulary of 200 from ~\cite{ref:durflex}. The speaker code $s$ is a learnable embedding over the ten ESD speakers. The CFM uses a $12$-layer DiT of dimension $512$, $8$ attention heads, and MLP expansion ratio $4$, while all acoustic models operate on $80$-dimensional mel-spectrograms at hop size $256$ and a $1024$-point FFT. We optimize all modules with AdamW~($\beta_1{=}0.9$, $\beta_2{=}0.98$) at learning rate $1\!\times\!10^{-4}$, with dropout $0.1$ in the DiT decoder and $0.3$ in the EVA. Prime-EVC is trained on a single NVIDIA A6000 GPU with batch size $16$ for $500$K iterations. RTF is measured on the same GPU at batch size $1$, covering the acoustic model and the BigVGAN vocoder.
Speech samples are available online\footnote{\url{https://yupan0v0.github.io/prime-evc/}}.

\section{Results and Analysis}
\label{sec:exp}

\subsection{Main Results}
\label{sec:main}

As shown in Table~\ref{tab:main}, the conventional GAN, sequence-to-sequence, and intensity-controllable systems (StarGAN-EVC, Seq2seq-EVC, Emovox) trail in naturalness and emotion conversion (UTMOS $\le 2.11$, EECS $\le 0.49$), while Textless-EVC and DurFlex-EVC form a stronger tier that still trails Prime-EVC. DurFlex-EVC is competitive only with multi-step sampling, approaching full fidelity at $100$ steps (UTMOS $3.21$, RTF $1.917$) but collapsing under single-step decoding (UTMOS $1.32$, EECS $0.32$). 
By contrast, with a \emph{single} Euler step Prime-EVC delivers the best objective quality on all evaluated axes (UTMOS $3.37$, CER $4.40\%$, WER $9.61\%$, EECS $0.87$, SECS $0.80$), cutting the remaining speaker-similarity gap to the vocoded ground truth from $0.08$ to $0.04$, while running $3.4$--$50\times$ faster (RTF $0.038$ vs.\ DurFlex-EVC's $0.132$--$1.917$). As Secs.~\ref{sec:ablation} and~\ref{sec:nfe} show, this margin stems from constructing an emotion-rich conditioning before generation, so a single step converts from an already-emotional target rather than reconstructing expressiveness during sampling. The subjective ratings corroborate this: Prime-EVC earns the highest nMOS ($3.77$) and eMOS ($3.90$), ahead of the $100$-step DurFlex-EVC ($3.69$/$3.83$).

\subsection{Ablation Study}

\label{sec:ablation}

\begin{table}[ht]
\centering
\caption{Ablation of Prime-EVC. Above the rule, each row removes one conditioning component from the full model. Below the rule are the conditioning alone and alternative decoder objectives.}
\label{tab:ablation}
\resizebox{\columnwidth}{!}{%
\renewcommand{\arraystretch}{1}
\begin{tabular}{lcccc}
\toprule
Model & WER$\downarrow$ & UTMOS$\uparrow$ & EECS$\uparrow$ & SECS$\uparrow$ \\
\midrule
\textbf{Prime-EVC} & 9.61 & 3.37 & 0.87 & 0.80 \\
\quad w/o FiLM        & 9.86 & 3.34 & 0.83 & 0.77 \\
\quad w/o intensity   & 9.55 & 3.35 & 0.82 & 0.80 \\
\quad w/o classifier  & 9.52 & 3.41 & 0.81 & 0.81 \\
\quad w/o STFT        & 9.17 & 3.23 & 0.85 & 0.80 \\
\quad w/o EVA         & 8.25 & 2.89 & 0.46 & 0.78 \\
\midrule
\quad $\mu$ only (no decoder) & 196.4 & 1.48 & -- & -- \\
\quad w/o CFM (regression) & 9.92 & 2.79 & 0.67 & 0.78 \\
\quad OT-CFM$\to$MeanFlow & 13.76 & 1.89 & 0.77 & 0.73 \\
\bottomrule
\end{tabular}}
\end{table}

As reported in Table~\ref{tab:ablation}, the EVA is decisive: its removal collapses EECS from $0.87$ to $0.46$ and UTMOS from $3.37$ to $2.89$, even close to the unenhanced-conditioning floor of the \textit{Global-cond.\ control} in Table~\ref{tab:nfe}.
The remaining terms are smaller and complementary: FiLM fusion, intensity scaling, and the classifier chiefly sharpen emotion (EECS $0.83$/$0.82$/$0.81\!\to\!0.87$, with FiLM also lifting SECS $0.77\!\!\to\!\!0.80$), while the spectro-temporal supervision improves naturalness (UTMOS $3.23\!\!\to\!\!3.37$) at a small but systematic intelligibility cost, as the full model favors emotion similarity over any single quality metric.
Below the rule, three variants bound the decoder's share of the work.
Vocoding $\mu$ directly yields unintelligible audio (UTMOS $1.48$, WER $>100\%$). A regression decoder (same DiT and conditioning) recovers only EECS $0.67$ (UTMOS $2.79$). This variant keeps the full conditioning, the EVA included, so the conditioning supplies the emotional content while the decoder objective governs how faithfully that content survives into the output. Finally, swapping OT-CFM for the mean-flow objective~\cite{ma2026meanvc} retains emotion (EECS $0.77$) but not quality (UTMOS $1.89$, WER $13.76\%$). 

\subsection{Effect of Sampling Steps}
\label{sec:nfe}
As illustrated in Table~\ref{tab:nfe}, we sweep the NFE for Prime-EVC and for a \textit{Global-cond.\ CFM} control that disables all conditioning enhancements, leaving the global style $c$ as the only emotion pathway.
The sweep confirms our design: Prime-EVC is already saturated at a single step, with all metrics essentially constant from one to sixteen steps while the RTF doubles ($0.038\!\to\!0.078$), making a single evaluation Pareto-optimal.
The control, in turn, isolates what extra steps actually buy.
It stays usable at a single step (UTMOS 2.80, WER 8.07), so single-step decoding is not itself the bottleneck. As the NFE grows, its signal fidelity improves (UTMOS $2.80\!\to\!3.24$), whereas the para-linguistic emotion (EECS $0.37\!\to\!0.44$) and the non-linguistic speaker identity (SECS $0.77\!\to\!0.78$) barely move.
Extra steps therefore buy signal fidelity rather than emotion or speaker identity, so the sampling budget is not the lever for expressive EVC.

\begin{table}[ht]
\centering
\caption{Effect of NFE for Prime-EVC and the \textit{Global-cond.\ CFM}, in which the global style is the only emotion pathway.}
\label{tab:nfe}
\renewcommand{\arraystretch}{0.9}
\resizebox{\columnwidth}{!}{%
\begin{tabular}{cccccc}
\toprule
NFE & UTMOS$\uparrow$ & WER(\%)$\downarrow$ & EECS$\uparrow$ & SECS$\uparrow$ & RTF$\downarrow$ \\
\midrule
\multicolumn{6}{l}{\textit{Prime-EVC}}\\
1  & 3.37 & 9.61 & 0.87 & 0.80 & 0.038 \\
4  & 3.35 & 9.70 & 0.88 & 0.80 & 0.045 \\
16 & 3.37 & 9.64 & 0.87 & 0.80 & 0.078 \\
\midrule
\multicolumn{6}{l}{\textit{Global-cond.\ CFM}}\\
1  & 2.80 & 8.07 & 0.37 & 0.77 & 0.036 \\
4  & 3.23 & 7.95 & 0.43 & 0.78 & 0.044 \\
16 & 3.24 & 7.87 & 0.44 & 0.78 & 0.077 \\
\bottomrule
\end{tabular}}
\end{table}

\section{Conclusion}
\label{sec:conc}
In this paper, we present Prime-EVC, a one-step framework for expressive emotional voice conversion via emotion-primed flow matching. Concretely, Prime-EVC offloads emotion modeling onto a supervised, emotion-rich conditioning built before CFM decoding, so a single ODE step suffices for expressive conversion. Experiments on ESD show that Prime-EVC surpasses strong multi-step baselines in objective and subjective quality while running $3.4$$\times$ faster than DurFlex-EVC's 4-step setting, with a step-count analysis showing that expressiveness primarily tracks the conditioning rather than the number of ODE steps. 


\footnotesize
\vfill\pagebreak
\bibliographystyle{IEEEbib}
\bibliography{refs}

\end{document}